\documentclass[structabstract]{aa}

\usepackage{txfonts}
\usepackage{natbib}
\usepackage{amsbsy}
\usepackage{graphicx}

\newcommand{\micron}{$\mu$m}
\newcommand{\el}[1]{\mathrm{#1}}

\newcommand{\ten}[1]{$10^{#1}$}
\newcommand{\scit}[2]{$#1\times10^{#2}$}
\newcommand{\scim}[2]{#1\times10^{#2}}

\newcommand{\ps}{s$^{-1}$}

\newcommand{\pcs}{cm$^{-2}$}

\newcommand{\eq}[1]{Eq.\ (\ref{eq:#1})}

\newcommand{\fig}[1]{Fig.\ \ref{fig:#1}}
\newcommand{\figg}[1]{Figure \ref{fig:#1}}
\newcommand{\tb}[1]{Table \ref{tb:#1}}

\newcommand{\pder}[2]{\frac{\partial #1}{\partial #2}}

\newcommand{\okep}{\Omega_\mathrm{K}}
\newcommand{\tacc}{t_\el{acc}}
\newcommand{\cs}{c_\el{s}}
\newcommand{\w}{H$_2$O}

\defcitealias{visser09a}{V09}
\newcommand{\vddd}{\citetalias{visser09a}}
\defcitealias{dullemond06a}{DAW06}
\newcommand{\daw}{\citetalias{dullemond06a}}

\begin{document}

\title{Sub-Keplerian accretion onto circumstellar disks}
\titlerunning{Sub-Keplerian accretion onto circumstellar disks}

\author{
R. Visser \inst{1}
 \and C.P. Dullemond \inst{2}
}

\institute{Leiden Observatory, Leiden University, P.O. Box 9513, 2300 RA Leiden, The Netherlands \\
           \email{ruvisser@strw.leidenuniv.nl}
 \and Max-Planck-Institut f\"ur Astronomie, K\"onigstuhl 17, 69117 Heidelberg, Germany \\
           \email{dullemon@mpia.de}
}

\authorrunning{Visser \& Dullemond}

\date{Received $<$date$>$ / Accepted $<$date$>$}
%\date{Resubmitted version \today}

% ______________________________________________________________________

\abstract
{Models of the formation, evolution and photoevaporation of circumstellar disks are an essential ingredient in many theories of the formation of planetary systems. The ratio of disk mass over stellar mass in the circumstellar phase of a disk is for a large part determined by the angular momentum of the original cloud core from which the system was formed. While full 3D or 2D axisymmetric hydrodynamical models of accretion onto the disk automatically treat all aspects of angular momentum, this is not so trivial for 1D and semi-2D viscous disk models.} % context (optional)
{Since 1D and semi-2D disk models are still very useful for long-term evolutionary modelling of disks with relatively little numerical effort, we investigate how the 2D nature of accretion affects the formation and evolution of the disk in such models. A proper treatment of this problem requires a correction for the sub-Keplerian velocity at which accretion takes place.} % aims
{We develop an update of our semi-2D time-dependent disk evolution model to properly treat the effects of sub-Keplerian accretion. The new model also accounts for the effects of the vertical extent of the disk on the accretion streamlines from the envelope.} % methods
{The disks produced with the new method are smaller than those obtained previously, but their mass is mostly unchanged. The new disks are a few degrees warmer in the outer parts, so they contain less solid CO\@. Otherwise, the results for ices are unaffected. The 2D treatment of the accretion results in material accreting at larger radii, so a smaller fraction comes close enough to the star for amorphous silicates to be thermally annealed into crystalline form. The lower crystalline abundances thus predicted correspond more closely to observed abundances than did earlier model predictions. We argue that thermal annealing followed by radial mixing must be responsible for at least part of the observed crystalline material.} % results
{} % conclusions (optional)


\maketitle

% ______________________________________________________________________

\section{Introduction}
\label{sec:intro}
With the enormous increase in the amount of high-quality observational data of circumstellar disks in the last few years, a clear picture is emerging of how these objects evolve in time \citep{jorgensen07a,jorgensen09a,looney07a,lommen08a,sicilia09a}. They form during the collapse of a pre-stellar cloud core, undergo a number of accretion events (FU Orionis and EX Lupi outbursts), live for 3 to 10 Myr, and, shortly before they are destroyed, open up huge gaps visible in the dust continuum and sometimes also in gas lines \citep{dalessio05a,goto06a,brittain07a,ratzka07a,brown08a,pontoppidan08b}. These physical changes are echoed in the evolution of their chemical composition and dust properties. Pre-stellar cores contain mostly simple hydrides, radicals and other small molecules, largely frozen out onto the cold dust grains \citep{bergin97b,lee04a}. A fully formed disk is predicted to contain a much richer chemical mixture with a wide variety of complex organic molecules \citep{rodgers03a,aikawa08a}, although only simple organics have been observed so far \citep{lahuis06a,carr08a,salyk08a}. The dust by this time has grown from less than a micron to millimetres and centimetres, and part of it has evolved from an amorphous to a crystalline structure \citep{bouwman01a,bouwman08a,vanboekel05a,natta07a,lommen07a,lommen09a,watson09a,olofsson09a}. Crystalline silicate dust is observed down to temperatures of 100 K, well below the threshold of 800 K required to convert amorphous silicates into crystalline form. One of the central questions of this paper is how much silicate material comes close enough to the star to be crystallised. We also investigate how the crystalline silicates end up so far outside of the hot inner disk where they appear to be formed.

One way to answer these questions is to construct detailed models of the evolution of circumstellar disks based on our current understanding of the physics of these objects, and then compare to the available observational data. However, it would require extraordinarily heavy computations to run a model that does justice to all physical processes known to be involved. A circumstellar disk ranges from a few stellar radii to hundreds of AU, and lasts for several million years. A full model would therefore have to resolve hundreds of millions of inner orbits, and span some five orders of magnitude on a spatial scale. Moreover, an accurate radiative transfer method is required to properly compute the temperatures. All this is clearly too demanding. Most multidimensional hydrodynamical simulations therefore solve sub-problems that only capture part of the disk, or only evolve over a limited time. Even these models, though, require days or weeks of CPU time for a single set of parameters.

An alternative approach is to parameterise most of the physics in some form, and treat the disk evolution as a simpler one-dimensional (1D) time-dependent problem. One assumes axisymmetry and integrates the density vertically to obtain the surface density $\Sigma$, which is now only a function of the radial coordinate $R$ and the time $t$. These kinds of models go back to the pioneering work by \citet{shakura73a} and \citet{lyndenbell74a}. In order to use these models throughout the disk's lifetime, some way must be found to also include the birth phase of the disk in a reasonably realistic way. Two-dimensional axisymmetric hydrodynamical models of disk formation show the presence of a stand-off shock that decelerates the supersonically infalling matter as it approaches the disk from above and below \citep[e.g.,][]{tscharnuter87a,yorke93a,neufeld94a}. The structure of this stand-off shock is clearly multidimensional in the outer regions, but may be approximated in a simpler manner in the inner regions \citep{nakamoto94a}. \citet{hueso05a} constructed a 1D disk evolution model with such an approximation and used it to analyse two T Tauri stars. This showed that simple models of disk formation and evolution can be very powerful and yield valuable insight into the evolutionary stage of young stellar objects. Similar models have also been used to analyse the statistics of the accretion rates measured in pre--main-sequence stars \citep{dullemond06c,vorobyov08a}.

Another problem addressed with these 1D parameterised models is the origin, evolution and transport of gas and dust in circumstellar disks. For instance, \citet[hereafter DAW06]{dullemond06a} suggested that the initial outward expansion of the disk during the disk formation phase (observationally the Class 0/I phase) may be very effective in transporting thermally processed dust to the outer parts of the disk. Based on that work, one would expect to find a number of disks with nearly 100\% crystalline dust. However, no such extremely crystalline disks are observed \citep{bouwman08a,watson09a}. This is one of the issues addressed in this paper.

Yet another application of parameterised disk evolution models was shown in \citet[hereafter V09]{visser09a}. We followed the envelope material from thousands of AU inwards, through the accretion shock and into the disk, to analyse when ices evaporate from and recondense onto the grains. Since much of the interesting physics happens in the outer regions of the disk (several hundred AU), where the accretion shock no longer has a simple 1D shape, some way had to be found to include the 2D axisymmetric nature of this region without having to resort to a full-scale multidimensional hydrodynamical simulation. Our recipe was to construct a semi-2D disk model, i.e., to generate the 2D density structure $\rho(R,z,t)$ (where $z$ is the vertical coordinate away from the midplane) out of the 1D surface density $\Sigma(R,t)$ using the disk temperature to compute the scale height at every radius. The accretion onto the disk was then modelled by following infalling matter on ballistic supersonic trajectories until its density equals the density of the disk. This yields a 2D axisymmetric shape of the stand-off shock that is similar to that obtained by full 2D axisymmetric hydrodynamical models \citep{yorke93a,brinch08a}. However, it is not trivial to link this type of multidimensional infall structure back to the 1D disk evolution model. The main obstacle is that a substantial fraction of the matter does not really fall onto the disk's {\em surface} but onto the disk's {\em outer edge}, i.e., it accretes from the side instead of from the top. Moreover, this matter rotates with a very sub-Keplerian velocity. If this is not treated in some way, the total angular momentum balance of the system is violated, potentially leading to very wrong estimates of the evolution of the disk mass with time. We devised an ad-hoc solution to this problem in \vddd{}, which gave reasonable results but did not properly conserve angular momentum. We derive a more rigorous solution in the current paper.

The problem of sub-Keplerian accretion onto a disk is not studied here for the first time. It has long been known that material falling in an elliptic orbit onto the surface of the disk has sub-Keplerian angular momentum, even in the case of infinitely flat disks. When mixing with the disk material, which is in Keplerian rotation, a torque is exerted that pushes the disk material in towards the star. \citet{cassen81a} described this problem and solved it elegantly. For disks without internal torque and with small vertical extent, they found an analytical solution for the disk surface density as a function of time, and they also presented numerical results for viscous disks. \citet{hueso05a} derived an alternative solution: they simply calculated the radius for which the specific angular momentum of the infalling matter is Keplerian, and inserted the matter into the disk at that radius. This is not unreasonable, because one may assume that the material, after it hits the surface of the disk, first adjusts its own orbit before it mixes with the disk material below. It may be argued that as long as the angular momentum budget is not violated, the disk evolution is not much affected by the precise treatment. However, if one wishes to follow the radial motion and mixing of certain chemical species or types of dust, this may depend more sensitively on the way the angular momentum problem is treated.

In this paper we add a consistent treatment of sub-Keplerian 2D accretion to a semi-2D model for the formation and viscous evolution of a disk (Sect.\ \ref{sec:eqs}). Some basic disk properties arising from the new treatment are discussed in Sect.\ \ref{sec:diskprop}, and the results from \vddd{} are briefly revisited in Sect.\ \ref{sec:gasice}. The dust crystallisation results from \daw{} are re-evaluated in Sect.\ \ref{sec:cryst}, and finally conclusions are drawn in Sect.\ \ref{sec:conc}.

% ______________________________________________________________________

\section{Equations}
\label{sec:eqs}
The model describes the formation and evolution of a circumstellar disk inside a collapsing cloud core. The disk's surface density $\Sigma$ is governed by two processes: accretion of material from the envelope and viscosity-driven diffusion. This requires the continuity equation
\begin{equation}
\label{eq:cont}
\pder{(\Sigma R)}{t} + \pder{(\Sigma Ru_R)}{R} = RS\,,
\end{equation}
with $u_R$ being the radial velocity and $S$ the source function that accounts for accretion from the envelope. In addition to conservation of mass, ensured by \eq{cont}, there must be conservation of angular momentum \citep{lyndenbell74a}:
\begin{equation}
\label{eq:cam}
\pder{(\Sigma\okep R^3)}{t} + \pder{(\Sigma\okep R^3u_R)}{R} = \pder{}{R}\left(\Sigma\nu R^3\pder{\okep}{R}\right) + S\Omega R^3\,,
\end{equation}
with $\nu$ being the viscosity coefficient (discussed further down) and $\okep$ the Keplerian rotation rate.

In the last term in \eq{cam}, belonging to the accreting material, the rotation rate $\Omega$ is smaller than $\okep$. If accretion {\em would} occur with the Keplerian velocity, \eq{cam} can be solved to give the expression for $u_R$ used by \daw:
\begin{equation}
\label{eq:veldaw}
u_R(R,t) = -\frac{3}{\Sigma\sqrt{R}}\pder{}{R}\left(\Sigma\nu\sqrt{R}\right)\,.
\end{equation}
Working from that solution, two methods have been used in the recent literature to correct for the sub-Keplerian velocity of the accreting material. \citet{hueso05a} and \daw{} modified the source function so that all incoming material accretes at the exact radius where its angular momentum equals that of the disk. One might picture this as a quick redistribution of material in the top layers of the disk after accretion initially occurred in a sub-Keplerian manner. A disadvantage of this method is that it introduces a discontinuity in the infall trajectories: upon accretion onto the disk, material instantaneously jumps to a smaller radius. Hence, in \vddd{} we chose to modify the expression for the radial velocity instead of that for the source function, taking
\begin{equation}
\label{eq:velold}
u_R(R,t) = -\frac{3}{\Sigma\sqrt{R}}\pder{}{R}\left(\Sigma\nu\sqrt{R}\right) - \eta_\el{r}\sqrt{\frac{GM_\ast}{R}}\,,
\end{equation}
with $M_\ast$ being the stellar mass at time $t$ and $\eta_\el{r}$ a parameter with a constant value of 0.002. While this method has the desired effect of transporting material to smaller radii without discontinuities, it does not properly conserve angular momentum.

As an alternative to these two methods, \eq{cam} can also be solved more generally for the case that $\Omega<\okep$, giving
\begin{equation}
\label{eq:velnew}
u_R(R,t) = -\frac{3}{\Sigma\sqrt{R}}\pder{}{R}\left(\Sigma\nu\sqrt{R}\right) - \frac{2RS}{\Sigma}\frac{\okep-\Omega}{\okep}\,.
\end{equation}
Expressions for $\Omega$ (or, rather, for the azimuthal velocity) may be found in \citet{cassen81a} and \citet{terebey84a}. The source function is taken from \vddd:
\begin{equation}
\label{eq:source}
S(R,t) = 2\rho u_z\,,
\end{equation}
with $\rho$ being the density at the disk-envelope boundary and $u_z$ the vertical component of the envelope velocity field at that point. Equations (\ref{eq:cont}) and (\ref{eq:velnew}) now describe a fully continuous solution to the evolution of the surface density with proper conservation of angular momentum. A physical interpretation of \eq{velnew} is provided in Sect.\ \ref{sec:diskprop}. Note that the model contains no prescription for photoevaporation or other disk dispersal mechanisms that come into play in the T Tauri or Herbig Ae/Be phase. This could easily be added \citep[e.g.,][]{alexander06b}, but we choose to focus only on the infall phase of star formation. Hence, our simulations are only reliable up to an age of about 1 Myr.

The boundary between the disk and the envelope was computed in \vddd{} as the surface where the density from both was the same. Here, we take instead the surface where the ram pressure of the infalling gas equals the thermal gas pressure from the disk:
\begin{equation}
\label{eq:pres}
\rho_\el{env}u^2 = \frac{kT_\el{s}\rho_\el{disk}}{\mu m_\el{p}}\,
\end{equation}
with $u$ being the velocity of the infalling material and $\mu$ the mean molecular mass of 2.3 proton masses per gas particle. The temperature at the surface of the disk, $T_\el{s}$, is determined by irradiation from the star and by viscous heating. The use of the isobaric surface as the disk-envelope boundary instead of the isopycnic surface can have some effect on the model results, but, as shown in the next sections, the effects of the new treatment of the sub-Keplerian accretion are usually much larger.

The viscosity coefficient appearing in Eqs.\ (\ref{eq:cam}) and (\ref{eq:velnew}) is calculated in the same way as in \daw{} and \vddd, i.e., with the $\alpha$ prescription from \citet{shakura73a}:
\begin{equation}
\label{eq:viscco}
\nu(R,t) = \frac{\alpha kT_\el{m}}{\mu m_\el{p}\okep}\,,
\end{equation}
with $T_\el{m}$ being the midplane temperature. Effects of self-gravity of the disk are ignored in \eq{viscco}, likely leading to an overestimate of the disk-to-star mass ratio for the more massive disks ($M_\el{d}\ga0.3M_\ast$) covered by our choice of initial conditions \citep{vorobyov10a}. For the current work, the $\alpha$ parameter is set to a constant value of 0.01 \citep{hartmann98a}. Choosing a different value, or allowing $\alpha$ to vary in time or with radius, would affect the evolution of the disk \citep[e.g.,][]{lin90a,kratter08a,vorobyov09a,rice09a,rice10a}. In order to keep the current work focused on the new treatment of the sub-Keplerian accretion, the effects of different viscosity prescriptions will be explored separately in a future publication.

% ______________________________________________________________________

\section{Size, mass and structure of the disk}
\label{sec:diskprop}
The net effect of the rightmost terms in Eqs.\ (\ref{eq:velold}) and (\ref{eq:velnew}) is the same: to provide an additional inward flux of matter. However, the radial dependence of the additional flux is different. In the old method (\eq{velold}), it was largest at the inner edge of the disk because of the $R^{-1/2}$ proportionality. In the new method, on the other hand, it is strongest at the {\em outer} edge of the disk because of the sharp decrease in surface density at that point. Physically, the new dependence is easy to understand. Near the disk's inner edge, accretion occurs into a large column of material (large $\Sigma$), and the torque arising from the different azimuthal velocities only results in a weak inward push. The same torque provides a much stronger push at the outer edge, where there is less disk material per unit surface (small $\Sigma$). In effect, the material falling onto the disk's outer edge pushes the disk inwards and limits its radial growth.

\begin{figure}[t!]
\resizebox{\hsize}{!}{\includegraphics{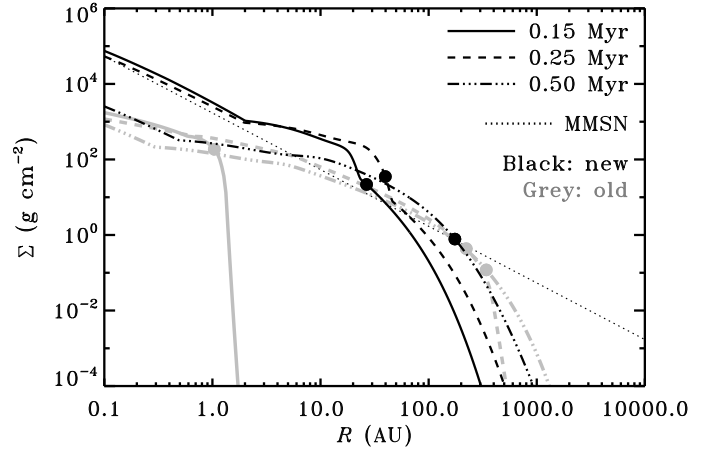}}
\caption{Surface density of the disk for the standard model from \vddd{} at 0.15, 0.25 and 0.50 Myr (0.6, 1.0 and 2.0 $\tacc$). The grey lines show the results from \vddd; the black lines show our new results. The filled circles indicate the radius that contains 90\% of the disk's mass. The thin dotted line is the minimum-mass solar nebula from \citet{hayashi81a}.}
\label{fig:oldnewsig}
\end{figure}

In order to quantify the size of the disk, we first have to define the outer edge $R_\el{d}$. Since the equations allow an infinitesimal part of the disk to spread to an infinitely large distance, we cannot simply use the full radial extent of the disk as its outer edge. Instead, we define $R_\el{d}$ as the radius that contains 90\% of the disk's mass. In general, the surface density at that point remains well above 0.01 g \pcs{} (see, e.g., \fig{oldnewsig}), so we do not have to worry about the outer parts being rapidly photoevaporated or dispersed for the duration of our simulations.

With the new method, the disk's outer edge initially lies beyond the centrifugal radius,
\begin{equation}
\label{eq:rcen}
R_\el{c}(t) = \frac{1}{16}\cs m_0^3t^3\Omega_0^2\,,
\end{equation}
where $\cs$ is the sound speed, $m_0$ is a numerical factor equal to 0.975 \citep{shu77a}, and $\Omega_0$ is the solid-body rotation rate. The outer edge grows more or less linearly in time as the disk spreads out to conserve the angular momentum of the entire system. The centrifugal radius grows as $t^3$, so the ratio $R_\el{d}/R_\el{c}$ decreases as the collapse goes on. If the rotation rate is high enough, the centrifugal radius overtakes the outer edge of the disk before the end of the collapse. For the remainder of the collapse phase, $R_\el{d}$ equals $R_\el{c}$, because the disk cannot be smaller than the centrifugal radius \citep{cassen81a}. If the rotation rate is low enough that the centrifugal radius does not overtake the outer edge of the disk before the end of the collapse, $R_\el{d}$ continues its near-linear growth until $\tacc$. After accretion from the envelope has ceased and the inward push from the accreting material is no longer present, the disk is free to expand more rapidly. Its outer edge again grows linearly in time in both the high- and low-rotation cases, and does so at a higher rate than in the initial linear-growth regime.

\begin{figure}
\resizebox{\hsize}{!}{\includegraphics{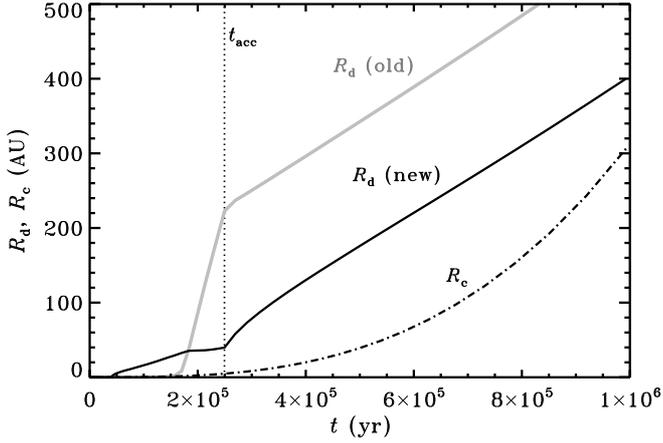}}
\caption{Outer disk radius as a function of time for the standard model from \vddd. Solid black: method from the current paper; grey: method from \vddd. The dash-dotted curve shows the centrifugal radius, and the dotted vertical line indicates the end of the envelope accretion phase. Note that $R_\el{c}$ is a physical quantity only up to $\tacc$; for larger $t$, we simply plot the same mathematical function (\eq{rcen}).}
\label{fig:rdisk}
\end{figure}

\figg{rdisk} shows the outer edge as a function of time for the standard model from \vddd, which has an initial core mass $M_0$ of 1.0 $M_\odot$, a sound speed $\cs$ of $0.26$ km \ps, and a low rotation rate $\Omega_0$ of \ten{-14} \ps. The disk grows to about 40 AU at the end of the collapse phase at $\tacc=\scim{2.5}{5}$ yr, and it spreads to ten times that size over the next \scit{7.5}{5} yr. The method from \vddd{} (labelled ``old'' in \fig{rdisk}) provides a more rapid growth during the collapse phase, giving an $R_\el{d}$ of 230 AU at $\tacc$. The post-infall spreading occurs at the same rate as in the new method. We find the same qualitative differences and similarities between the old and the new method for the rest of the parameter grid from \vddd.

Is the smaller disk from the new method a realistic scenario? Our model necessarily assumes $u_R$ to be the same at all heights above the midplane. In reality, one might expect the infalling material to interact mostly with the surface of the disk, exerting its torque on only a fraction of the total surface density. This pushes the surface material radially inwards, while material near the midplane is unaffected. Indeed, the hydrodynamical simulations of \citet{brinch08a} show such behaviour, with material near the surface moving inwards and material at the midplane moving outwards. However, as this midplane material moves out beyond the outer edge, it becomes surface material itself, and is in turn exposed to the inward push from the envelope material. The assumed altitude-independence of $u_R$ may therefore lead to the outer radius being underestimated by a factor of two or three. On the other hand, the models of \citet{rice09a} and \citet{rice10a} suggest that disks with a spatially dependent viscous $\alpha$ expand less slowly than do disks with a constant $\alpha$, in which case we would be overestimating the outer radius of the disk. Altogether, the new method produces disk sizes that are consistent with observations and other simulations, and it offers an improvement over the unrealistically large disks sometimes produced with the old method.

Although the disks are smaller with the new method, they are not necessarily less massive. In fact, we find a rather large mass increase for the standard model from \vddd: from 0.05 to 0.13 $M_\odot$ at the end of the collapse phase. There are three causes for this, each of which accounts for 0.02--0.03 $M_\odot$: (1) the new definition of the disk-envelope boundary (\eq{pres}); (2) the new sub-Keplerian correction (\eq{velnew}); and (3) an improved integration scheme in the computational code. The effects are smaller for more rapidly rotating clouds. For example, the disk mass obtained for the reference model from \vddd{} ($M_0=1.0$ $M_\odot$, $\cs=0.26$ km \ps, $\Omega_0=10^{-13}$ \ps) is unchanged at 0.43 $M_\odot$. Observed disk masses are typically between 0.001 and 0.1 $M_\odot$ for low-mass protostars \citep[e.g.][]{andrews07b,andrews07a}. For realistic initial conditions (i.e., $\Omega_0=10^{-14}$ \ps{} rather than \ten{-13} \ps), our model is known to overpredict the disk-to-star mass ratio by about a factor of three compared to observations \citep{jorgensen09a}. However, it has also been argued that observed disk masses have been systematically underestimated \citep{hartmann06a}, a view supported by the hydrodynamical simulations of \citet{vorobyov09b} and \citet{kratter10a}.

The combination of smaller sizes and equal or larger masses means that the surface density profiles have also changed from the old to the new method. \figg{oldnewsig} shows the surface density for the standard model from \vddd{} at three different time steps. The time evolution is qualitatively the same for the old (grey) and the new method (black). Up to the end of the accretion phase ($t=\tacc$; dashed curves), mass is mostly added onto the outer parts of the disk, and the surface density inside of $R_\el{d}$ remains nearly constant. At later times, when there is no more accretion from the envelope, the disk continues to spread and the surface density inside of $R_\el{d}$ gradually decreases. Quantitatively, the rate at which the surface density changes is different for the old and the new method, and also depends on the choice of initial conditions.

At $\tacc$, the surface density from our new method falls off as $R^{-1.5}$ inside of 2 AU. This is the same slope as in the original model for the minimum-mass solar nebula \citep[MMSN;][]{weidenschilling77b,hayashi81a}, but it is much shallower than the slope of $-2.2$ in the new MMSN model of \citet{desch07a}. The slope of $-1.5$ is also commonly found in other disk evolution simulations \citep[e.g.,][]{vorobyov09a,rice10a}. However, whereas those simulations tend to have a constant slope out to at least a few tens of AU, our surface density flattens off to a slope of $-0.7$ between 2 and 30 AU. This is due to the mass being accreted mostly onto the outer parts of the disk. Indeed, the old method, where the infalling material was distributed more evenly across the entire disk, shows a somewhat steeper slope of about $-1.0$ between 2 and 30 AU.

% ______________________________________________________________________

\section{Gas-ice ratios}
\label{sec:gasice}
During the collapse of the pre-stellar core to form a protostar and circumstellar disk, large changes occur in both density and temperature. Many molecular species are frozen out onto dust grains before the onset of collapse \citep{bergin97b,lee04a}. The warm-up phase during the collapse causes some of them to evaporate, and they may freeze out again once material settles near the disk's relatively cold midplane. In \vddd{}, we modelled these processes for carbon monoxide (CO) and water (\w). Here, we investigate whether our new sub-Keplerian accretion correction affects these results.

The model consists of several steps. First, it computes the 2D axisymmetric density and velocity structure at regular intervals from the onset of collapse ($t=0$) to the point where the entire envelope has accreted onto the star and disk ($t=\tacc$). Dust temperatures are computed at the same time intervals with the radiative transfer code RADMC \citep{dullemond04a}, and the gas temperature is assumed equal to the dust temperature. Using the velocity structure, the model then computes infall trajectories from a grid of initial positions in the envelope. Each trajectory represents an individual parcel of gas and dust, for which we now know the density and temperature as a function of time and position. This allows us to compute the adsorption and desorption rates of CO and \w, and solve for their gas and ice abundances in a Lagrangian frame. Finally, the parcels' abundances are transformed back into 2D axisymmetric abundance profiles for the disk and remnant envelope.

As discussed in Sect.\ \ref{sec:diskprop}, the main difference between the disk properties from \vddd{} and the new method is the size of the disk. Because the mass has increased or stayed the same (depending on the model parameters), the density of the disk is now higher. In the absence of viscous and shock heating (see below), this leads to a more rapid decrease of the dust temperature along the midplane at small radii. For example, at $t=\tacc$ in the reference model from \vddd{} ($M_0=1.0$ $M_\odot$, $\cs=0.26$ km \ps, $\Omega_0=10^{-13}$ \ps), the temperature at 30 AU is 45 K\@. With the new method, the temperature is down to 31 K at that point. The midplane temperature decreases further with radius until it reaches 21 K at 160 AU with the new method. At larger radii, photons scattering off the surface of the disk begin to reach the midplane again and the temperature gradually increases to 25 K at the outer edge at 500 AU. The original disk in \vddd{} extended to 1500 AU, and the turnover from a decreasing to an increasing temperature along the midplane occurred at 1000 AU instead of at 160 AU. Hence, even though the midplane temperature decreased less rapidly in \vddd{}, it still reached a lower minimum: 15 K, as opposed to 21 K with the new method.

If all CO is taken to desorb at 18 K, as it would for a pure CO ice, the new temperature profile does not allow for any solid CO to exist in this particular model at $\tacc$. At later times, when the disk has spread to larger sizes and the protostar has become less luminous (\vddd), a region appears around the midplane where the temperature does go below 18 K and CO freezes out again. In reality, solid CO forms a mixture with solid \w, and some of the CO remains trapped in the ice matrix at temperatures above 18 K (\citealt{collings04a}; \citealt{viti04a}; Fayolle et al.\ in prep.). When the gas-ice ratios are computed accordingly, the mass fraction of solid CO averaged over the entire disk at $\tacc$ goes from 33\% with the old method to 20\% with the new method.

The gas-ice ratios for pure CO in the standard model from \vddd{} ($M_0=1.0$ $M_\odot$, $\cs=0.26$ km \ps, $\Omega_0=10^{-14}$ \ps) are unchanged. With both the old and the new method, the entire disk is warmer than 18 K at $\tacc$, so there is no solid CO\@. If we allow part of the CO to be trapped in the \w{} ice, the disk-averaged solid fraction is 15\% with both methods.

Massive disks ($M_\el{d}>0.3M_\ast$) are susceptible to heating sources other than the stellar radiation field, such as viscosity and gravitational shocks. These heating mechanisms are not included in the chemical part of our computational code, where the gas-ice ratios are computed. (Viscous heating is, however, included in the part of the code that calculates the evolution of the mass and surface density of the disk and the evolution of the crystalline silicate abundances.) Viscous and shock heating would result in higher temperatures for the more compact disks produced by our new method than they would for the more extended disks produced by our old method. However, this is unlikely to change the solid fraction of CO by much. At the end of the collapse phase, the entire disk is already warmer than the evaporation temperature of pure CO (18 K), so all CO is predicted to be in the gas phase. Adding additional sources of heat will only strengthen that conclusion. For CO ice mixed with \w{} ice, the fraction of trapped CO hardly changes between the respective evaporation temperatures of CO and \w{} (Fayolle et al.\ in prep.), so the additional heating from viscosity and shocks would have to bring the temperature to more than 100 K to change the CO ice abundance significantly. This will only happen in a small zone around the snowline, so the mass fraction of solid CO averaged over the entire disk will remain close to the value of 20\% mentioned above.

In summary, the new treatment of the sub-Keplerian accretion results in disks that are a few degrees colder in the inner parts and a few degrees warmer in the outer parts. Overall, the new CO ice abundances are between 0 and 50\% smaller than those obtained with the old method. In all cases, \w{} remains solid in the entire disk except for the inner few AU.

% ______________________________________________________________________

\section{Crystalline silicates}
\label{sec:cryst}

% __________________________________________________

\subsection{Observations and previous model results}
\label{subsec:oldres}
Infrared spectroscopic observations have shown that 1--30\% of the silicate dust in the disks around Herbig Ae/Be and classical T Tauri stars occurs in crystalline form \citep{bouwman01a,bouwman08a,vanboekel05a}. Even larger fractions (up to 100\%) are found in the inner 1 AU of some sources \citep{watson09a}. The interstellar medium, from which this dust originates, has a crystalline fraction of at most 1--2\% \citep{kemper04a,kemper05a}. Two mechanisms are thought to dominate the conversion of amorphous silicates into crystalline form during the formation and evolution of a circumstellar disk. At temperatures above $\sim$1200 K, the original grains evaporate \citep{petaev05a}. When the gas cools down again, the silicates recondense in crystalline form \citep{davis03a,gail04a}. Alternatively, amorphous dust can be thermally annealed into crystalline dust at temperatures above $\sim$800 K \citep{wooden05a}. As the disk is formed out of the parent envelope, part of the infalling material accretes close enough to the star that it is heated to more than 800 or 1200 K and can be crystallised. However, the observations show significant fractions of crystalline dust at least out to radii corresponding to a temperature of 100 K\@. Crystalline dust is also found in comets, which are formed in regions much colder than 800 K \citep{wooden99a,keller06a}. This suggests an efficient radial mixing mechanism to transport crystalline material from the hot inner disk to the colder outer parts \citep{nuth99a,bockelee02a,keller04a}.

An argument against large-scale radial mixing was recently provided by spatially resolved observations with the Spitzer Space Telescope. \citet{bouwman08a} found a clear radial dependence in the relative abundances of forsterite and enstatite, two common specific forms of crystalline silicate. If both are formed in the hot inner disk and then transported outwards, one would expect the same relative abundances throughout the entire disk. Hence, the observed radial dependence argues in favour of a localised crystallisation mechanism such as heating by shock waves triggered by gravitational instabilities \citep{harker02a,desch05a}. However, the observations do not completely rule out the possibility of crystallisation in the hot inner disk followed by radial mixing; at best, they provide an upper limit to how much crystalline dust can be formed that way.

In another set of Spitzer observations, crystalline spectroscopic features in the 20--30 \micron{} region were detected three times more frequently than the crystalline feature at 11.3 \micron{} \citep{olofsson09a}. This is unexpected, because shorter wavelengths trace warmer material at shorter distances from the protostar, where all models predict the crystalline fractions to be larger. The crystalline 11.3 \micron{} feature may be partially shielded by the amorphous 10 \micron{} feature, but Olofsson et al.\ showed that this alone cannot explain the observations. A full compositional analysis (Olofsson et al.\ subm.) is required to shed more light on this ``crystallinity paradox''.

In the model of \daw, crystallisation occurs right from the time when the disk is first formed. Indeed, because the disk is very small at that time, its dust is hot and nearly fully crystalline. As the collapse proceeds and the disk's outer radius grows, an ever larger fraction of the infalling material does not come close enough to the star anymore to be heated above 800 K\@. In the absence of strong shocks, this results in amorphous dust being mixed in with the crystalline material. Hence, the crystalline fraction averaged over the entire disk is expected to decrease with time. There is tentative observational support for an age-crystallinity anticorrelation \citep{vanboekel05a,apai05a}, but this is far from conclusive \citep{bouwman08a,watson09a}. One should of course consider the fact that observations do not probe the entire disk. If the model results from \daw{} are interpreted over a limited part of the disk, such as the 10--20 AU region, they show only a small difference in the crystalline fractions at 1 and 3 Myr. Add to that the uncertainties in the ages for individual objects, and it is clear that the model results cannot be said to conflict the observational data.

The crystalline fractions obtained by \daw{} were all on the high end of the observed range of 1--30\%, unless unreasonably high initial rotation rates were adopted for the envelope or the disk temperature was lowered artificially. If we accept that only part of the silicates in the outer disk originate in the hot inner region -- so the other part, formed in situ, can account for the observed radial abundance variations -- the discrepancy between the \daw{} model and the observations becomes even larger. In the following, we show that we obtain more realistic crystalline fractions with our new method.

% __________________________________________________

\subsection{New model results}
\label{subsec:newres}
\begin{figure}[t!]
\resizebox{\hsize}{!}{\includegraphics{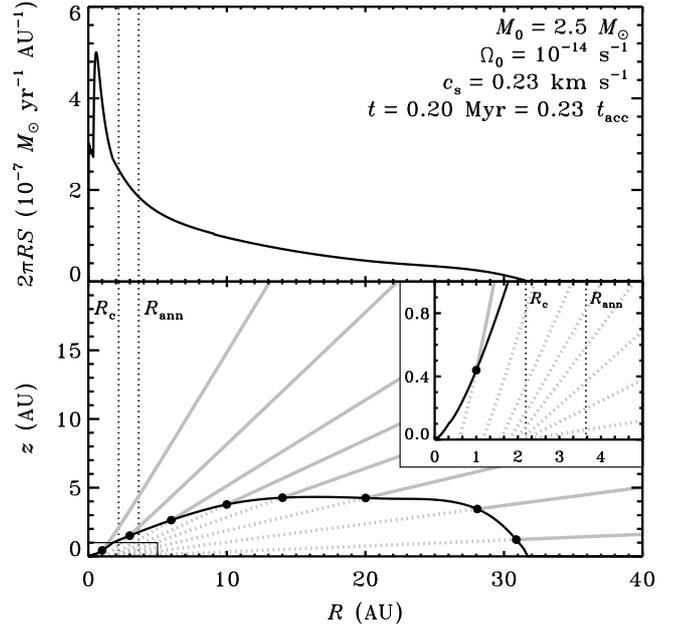}}
\caption{Accretion at 0.20 Myr (0.23 $\tacc$) for the default model of \daw. The vertical dotted lines indicate the current values of $R_\el{c}$ (2.2 AU) and $R_\el{ann}$ (3.6 AU). \textit{Top}: mass-loading as a function of radius. \textit{Bottom}: infall trajectories (solid grey) from the envelope onto the surface of the disk (black). In the absence of the disk, the trajectories would extend to the midplane along the dotted lines. The inset in the bottom panel shows a blow-up of the inner $5\times1$ AU.}
\label{fig:oldnewacc}
\end{figure}

The two main differences between the old method of \daw{} and our new method are (1) the treatment of the disk as a multidimensional object instead of just a flat accretion surface and (2) the improved solution to the problem of sub-Keplerian accretion (Sect.\ \ref{sec:eqs}). The former has the largest impact on the crystallisation. In case of a fully flat disk, material falls in along ballistic trajectories until it hits the midplane at or inside the centrifugal radius. If the vertical extent of the disk is taken into account, the infalling material hits the disk before it can flow all the way to the midplane. Because part of the disk often spreads beyond the centrifugal radius, especially at early times (\fig{rdisk}), accretion now occurs at much larger radii. This is visualised in \fig{oldnewacc}, which shows the mass loading onto the disk at \scit{2.0}{5} yr (0.23 $\tacc$) after the onset of collapse. The model parameters are those of the default model of \daw: an initial cloud core mass $M_0$ of 2.5 $M_\odot$, a rotation rate $\Omega_0$ of \scit{1}{-14} \ps, and a sound speed $\cs$ of 0.23 km \ps. The centrifugal radius at \scit{2.0}{5} yr is 2.2 AU, but the disk has already spread to 32 AU\@. Accretion occurs across the entire disk, although most mass falls in at small radii.

If the vertical structure of the disk is ignored when calculating the source function, as happened in the old method of \daw, the infall trajectories continue along the dotted lines. They all intersect the midplane inside of $R_\el{c}$. In this case, that means all accretion takes place inside the ``annealing radius'' ($R_\el{ann}$, the radius corresponding to 800 K) and all dust is turned into crystalline form. In the new method, only 36\% of the accreting material comes inside of $R_\el{ann}$, so the disk gains a much smaller amount of crystalline dust.

\begin{figure}[t!]
\resizebox{\hsize}{!}{\includegraphics{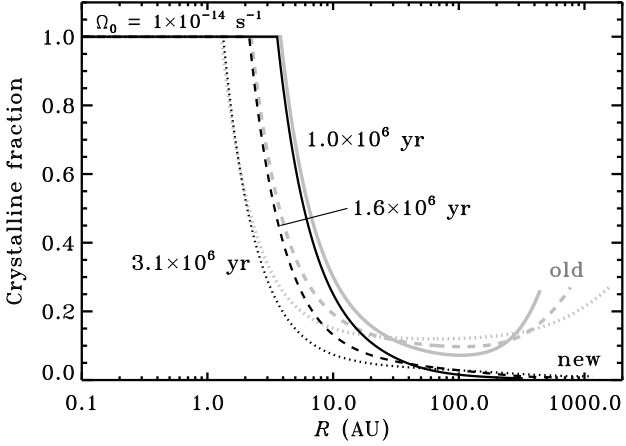}}
\caption{Fraction of crystalline silicates in the disk for the default model of \daw{} at 1.0 (solid), 1.6 (dashed) and 3.1 Myr (dotted). The grey lines show the results from \daw; the black lines show our new results. Lines are drawn up to $R_\el{d}$.}
\label{fig:cryst}
\end{figure}

The crystalline fractions obtained with the new method are compared to the results from \daw{} in \fig{cryst}. In both cases, the inner part of the disk, out to a few AU, is fully crystalline. This is followed by a near-powerlaw decrease as crystalline material is mixed to larger radii. At a few tens of AU, the crystallinity levels off to a base value that remains roughly constant to the outer edge of the disk, where the curves are terminated. The increase in crystallinity towards large radii in the original \daw{} curves is an artifact from the pre-infall low-density seed disk (used for reasons of numerical stability) and occurs only in those regions where the surface density is anyway nearly zero. This increase is therefore irrelevant. Due to the use of an improved numerical integration scheme, it is no longer present in the new curves.

\begin{figure}[t!]
\resizebox{\hsize}{!}{\includegraphics{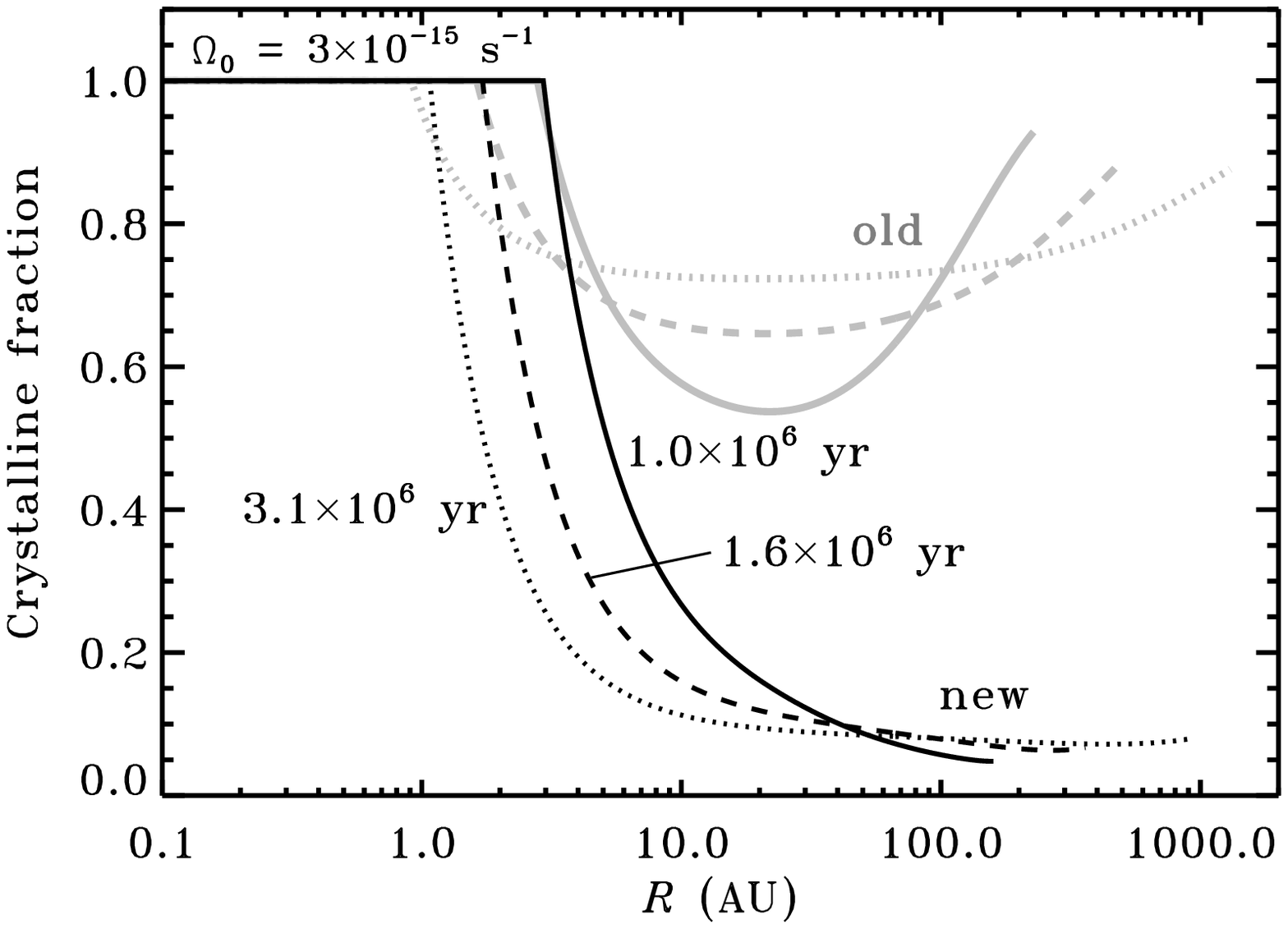}}
\caption{As \fig{cryst}, but for $\Omega_0=\scim{3}{-15}$ \ps.}
\label{fig:cryst3-15}
\end{figure}

At each of the three time steps plotted in \fig{cryst}, the new crystallinity outside of a few AU is lower than the old one. For example, at 10 AU and 3.1 Myr, the fraction is down from 15.1 to 7.4\%. At the outer edge of the disk, the new method produces crystalline fractions down to 1\%. The differences between the old and the new method become even more pronounced when we take a slightly lower initial rotation rate of \scit{3}{-15} \ps{} (\fig{cryst3-15}). The crystalline fraction at 10 AU and 3.1 Myr is now down from 72.6 to 11.3\%.

The amount of crystallisation that takes place during the formation and evolution of the disk depends on the initial conditions. \daw{} already discussed the effect of the rotation rate ($\Omega_0$) of the collapsing envelope. The more rapid the rotation, the larger the radius at which the bulk of the accretion takes place. This results in less material being heated above 800 K, so the disk becomes less crystalline. This effect also occurs in our new model. Two other conditions that can easily be changed are the initial cloud core mass ($M_0$) and the effective sound speed ($\cs$). In order to get a first understanding of their effects, we computed the crystalline fractions for the parameter grid from \vddd, as well as for models with initial masses of 0.2 and 5.0 $M_\odot$. \tb{pargrid} lists the relevant model parameters together with the fraction of crystalline silicates at three different positions and three different times. The first of the three positions (10 AU) is representative of the region probed by the recent Spitzer observations, while the other two positions (30 and 100 AU) contain colder material that can be studied with the Herschel Space Observatory. Results for models where the disk mass exceeds $\sim$30\% of the stellar mass should be treated with caution, because we do not include effects of self-gravity when computing the viscous evolution of the disk \citep{vorobyov10a}.

The models with a low sound speed all have a lower crystallinity than the models with a high sound speed. A lower sound speed gives a lower accretion rate from the envelope to the disk, as well as a lower accretion rate from the disk to the star. The net mass gain for the disk (accretion from envelope to disk minus accretion from disk to star) is also lower. However, this is more than offset by the accretion time becoming longer. At the end of the accretion phase, the disk mass is therefore higher for models with a low sound speed than it is for models with a high sound speed (\tb{pargrid}). The lower accretion rate in the low-$\cs$ models also gives a lower stellar luminosity, so the region where silicates can be crystallised is smaller. These effects combine to give smaller fractions of crystalline material throughout the entire disk.

\begin{table*}
\caption{Crystalline silicate fractions for a range of model parameters.}
\label{tb:pargrid}
\centering
\begin{tabular}{cccccccccccccccc}
\hline\hline
& \multicolumn{15}{c}{\rule{0pt}{1em}$\Omega_0=10^{-14}$ \ps, $\cs=0.19$ km \ps} \\
\hline
\rule{0pt}{1em}$t$ & \multicolumn{15}{c}{$R$ (AU)} \\
\cline{2-16}
\rule{0pt}{1em}(Myr) & 10 & 30 & 100 & & 10 & 30 & 100 & & 10 & 30 & 100 & & 10 & 30 & 100 \\
\hline
& \multicolumn{3}{c}{\rule{0pt}{1em}$M_0=0.20$, $M_\el{d}=0.012$} & & \multicolumn{3}{c}{\rule{0pt}{1em}$M_0=0.50$, $M_\el{d}=0.074$} & & \multicolumn{3}{c}{\rule{0pt}{1em}$M_0=1.0$, $M_\el{d}=0.26$} & & \multicolumn{3}{c}{\rule{0pt}{1em}$M_0=5.0$, $M_\el{d}=2.3$}\\
\hline
1.0 & 19.7 & 20.1 & 21.5 & & 6.7 & 6.0 & 5.5 & & 7.6 & 3.8 & 1.9 & & 17.0 & 2.9 & 0.4 \\
1.6 & 21.1 & 21.3 & 22.2 & & 6.1 & 5.8 & 5.7 & & 5.3 & 3.4 & 2.3 & & 23.0 & 3.9 & 0.4 \\
3.1 & 22.9 & 23.0 & 23.4 & & 5.9 & 5.8 & 5.9 & & 3.3 & 2.6 & 2.2 & & 24.6 & 4.4 & 0.6 \\
\hline\hline
& \multicolumn{15}{c}{\rule{0pt}{1em}$\Omega_0=10^{-14}$ \ps, $\cs=0.26$ km \ps} \\
\hline
\rule{0pt}{1em}$t$ & \multicolumn{15}{c}{$R$ (AU)} \\
\cline{2-16}
\rule{0pt}{1em}(Myr) & 10 & 30 & 100 & & 10 & 30 & 100 & & 10 & 30 & 100 & & 10 & 30 & 100 \\
\hline
& \multicolumn{3}{c}{\rule{0pt}{1em}$M_0=0.20$, $M_\el{d}=0$\,\tablefootmark{a}} & & \multicolumn{3}{c}{\rule{0pt}{1em}$M_0=0.50$, $M_\el{d}=0.037$} & & \multicolumn{3}{c}{\rule{0pt}{1em}$M_0=1.0$, $M_\el{d}=0.13$} & & \multicolumn{3}{c}{\rule{0pt}{1em}$M_0=5.0$, $M_\el{d}=1.9$}\\
\hline
1.0 & \ldots & \ldots & \ldots & & 32.1 & 32.3 & 33.8 & & 17.4 & 16.0 & 15.1 & & 54.1 & 9.3 & 1.2 \\
1.6 & \ldots & \ldots & \ldots & & 33.3 & 33.5 & 34.4 & & 16.4 & 15.7 & 15.4 & & 37.5 & 9.7 & 2.2 \\
3.1 & \ldots & \ldots & \ldots & & 35.1 & 35.3 & 35.7 & & 16.0 & 15.8 & 15.8 & & 16.5 & 6.1 & 3.3 \\
\hline\hline
& \multicolumn{15}{c}{\rule{0pt}{1em}$\Omega_0=10^{-13}$ \ps, $\cs=0.19$ km \ps} \\
\hline
\rule{0pt}{1em}$t$ & \multicolumn{15}{c}{$R$ (AU)} \\
\cline{2-16}
\rule{0pt}{1em}(Myr) & 10 & 30 & 100 & & 10 & 30 & 100 & & 10 & 30 & 100 & & 10 & 30 & 100 \\
\hline
& \multicolumn{3}{c}{\rule{0pt}{1em}$M_0=0.20$, $M_\el{d}=0.051$} & & \multicolumn{3}{c}{\rule{0pt}{1em}$M_0=0.50$, $M_\el{d}=0.23$} & & \multicolumn{3}{c}{\rule{0pt}{1em}$M_0=1.0$, $M_\el{d}=0.58$} & & \multicolumn{3}{c}{\rule{0pt}{1em}$M_0=5.0$, $M_\el{d}=4.1$}\\
\hline
1.0 & 2.1 & 1.9 & 1.6 & & 2.6 & 1.3 & 0.6 & & 4.4 & 1.3 & 0.3 & & 5.4 & 1.1 & 0.2 \\
1.6 & 1.8 & 1.7 & 1.6 & & 1.7 & 1.0 & 0.5 & & 2.9 & 1.1 & 0.4 & & 5.1 & 1.2 & 0.2 \\
3.1 & 1.6 & 1.5 & 1.5 & & 1.0 & 0.6 & 0.4 & & 1.5 & 0.6 & 0.3 & & 8.5 & 0.5 & \phantom{0}0.02 \\
\hline\hline
& \multicolumn{15}{c}{\rule{0pt}{1em}$\Omega_0=10^{-13}$ \ps, $\cs=0.26$ km \ps} \\
\hline
\rule{0pt}{1em}$t$ & \multicolumn{15}{c}{$R$ (AU)} \\
\cline{2-16}
\rule{0pt}{1em}(Myr) & 10 & 30 & 100 & & 10 & 30 & 100 & & 10 & 30 & 100 & & 10 & 30 & 100 \\
\hline
& \multicolumn{3}{c}{\rule{0pt}{1em}$M_0=0.20$, $M_\el{d}=0.019$} & & \multicolumn{3}{c}{\rule{0pt}{1em}$M_0=0.50$, $M_\el{d}=0.13$} & & \multicolumn{3}{c}{\rule{0pt}{1em}$M_0=1.0$, $M_\el{d}=0.43$} & & \multicolumn{3}{c}{\rule{0pt}{1em}$M_0=5.0$, $M_\el{d}=3.5$}\\
\hline
1.0 & 12.1 & 12.0 & 12.0 & & 5.3 & 4.4 & 3.6 & & 6.2 & 3.3 & 1.7 & & 19.3 & 3.9 & 0.6 \\
1.6 & 12.0 & 12.0 & 12.0 & & 4.5 & 4.0 & 3.6 & & 4.5 & 2.6 & 1.6 & & 14.6 & 3.3 & 0.7 \\
3.1 & 12.1 & 12.1 & 12.2 & & 3.7 & 3.5 & 3.3 & & 2.7 & 1.7 & 1.3 & & \phantom{0}8.6 & 2.4 & 0.8 \\
\hline
\end{tabular}
\tablefoot{
The fractions are given in per cent of the total silicate dust abundance at the indicated distance from the star ($R$) and time after the onset of collapse ($t$). The initial cloud core mass ($M_0$, in $M_\odot$) and the disk mass at the end of the accretion phase ($M_\el{d}$, in $M_\odot$) are listed for each combination of parameters.\\
\tablefoottext{a}{No disk is formed at all for this combination of parameters.}
}
\end{table*}

\begin{figure}
\resizebox{\hsize}{!}{\includegraphics{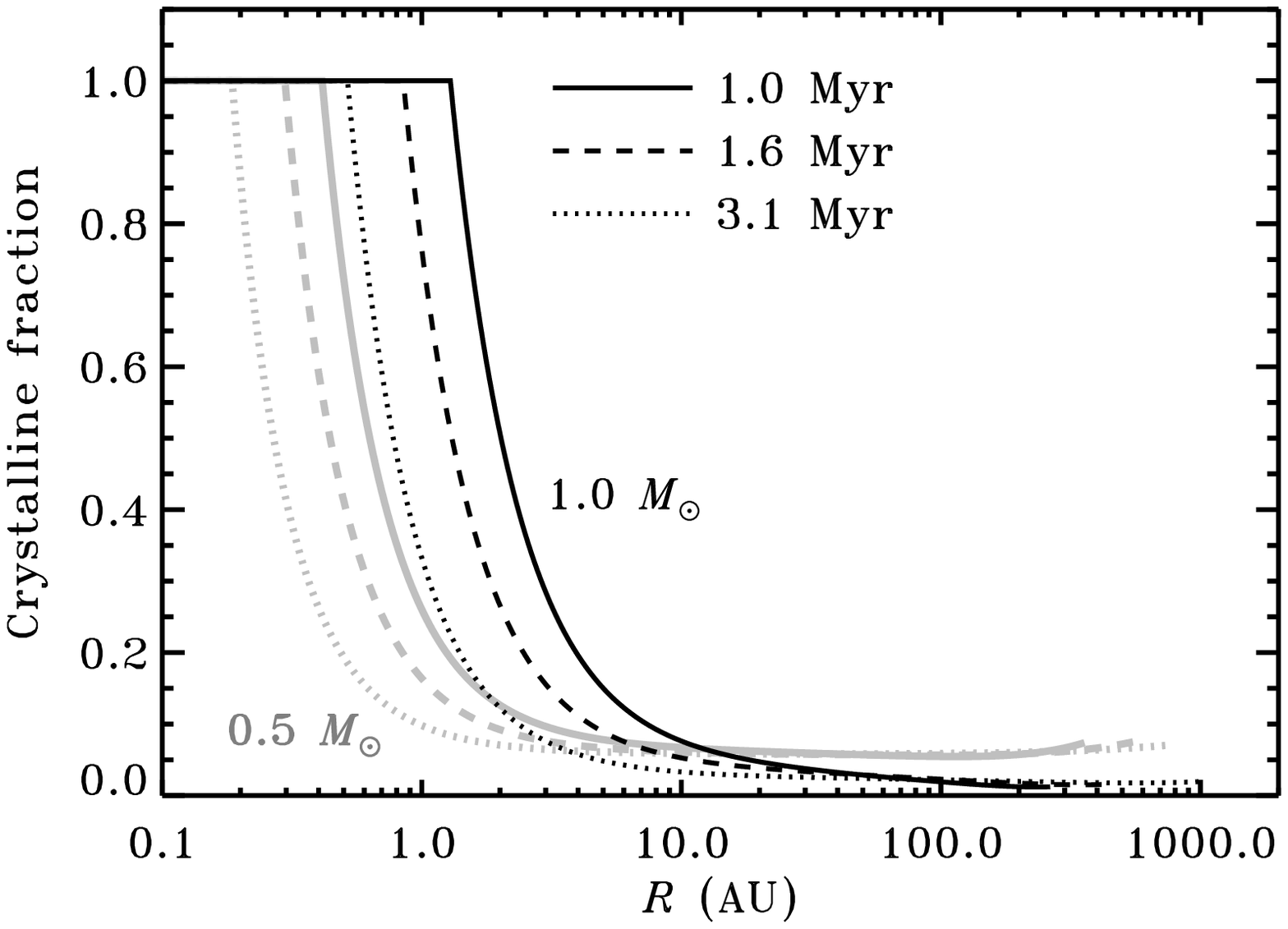}}
\caption{As \fig{cryst}, but for two of the models from \tb{pargrid} with $\Omega_0=10^{-14}$ \ps{} and $\cs=0.19$ km \ps. Grey: $M_0=0.5$ $M_\odot$; black: $M_0=1.0$ $M_\odot$.}
\label{fig:crystmass}
\end{figure}

Changing the initial mass of the cloud core has a more complicated effect on the crystallinity. \tb{pargrid} shows several cases where, for models that differ only in the initial mass, the crystalline fractions increase towards larger $M_0$, and several cases where they decrease in that direction. Due to the inside-out nature of the \citet{shu77a} collapse, models of different mass initially evolve in exactly the same way. However, the accretion phase of the higher-mass models in our grid lasts longer ($\tacc \propto M_0$) than that of the lower-mass models. This affects the crystallinity in two opposing ways. First, the protostar becomes more luminous for the higher-mass models \citep{dantona94a}, so the region in which crystallisation takes place is larger. Second, the disk grows larger, so the bulk of the accretion occurs farther from the star. The first effect results in higher crystalline abundances in the inner parts of the disk in the higher-mass models, while the second effect eventually results in \emph{lower} crystalline abundances in the \emph{outer} parts (\fig{crystmass}). Depending on the exact initial conditions, the transition from the inner to the outer disk in this context may lie inwards or outwards of the 10--100 AU region given in \tb{pargrid}, or even within that region.

Examples of two of these three possibilities can be found in the series of models with $\Omega_0=10^{-14}$ \ps{} and $\cs=0.19$ km \ps{} (top part of \tb{pargrid}). Going from $M_0=0.2$ to 0.5 $M_\odot$, the fraction of material accreting close enough to the star to be crystallised is reduced, so we find smaller crystalline fractions at 10, 30 and 100 AU: $\sim$6\% for the 0.5 $M_\odot$ model versus $\sim$20\% for the 0.2 $M_\odot$ model. Increasing the initial mass to 1.0 $M_\odot$, the higher stellar luminosity increases the crystallinity at 10 AU to 7.6\%. However, the crystallinity at larger radii suffers from the larger fraction of dust that remained amorphous during the accretion phase because the larger disk prevented it from getting closer to the protostar. At 30 AU, the crystalline fraction at 1.0 Myr decreases from 6.0 to 3.8\% when the initial mass goes from 0.5 to 1.0 $M_\odot$. The decrease in crystallinity at 100 AU is even larger: from 5.5 to 1.9\%. The differences in the other three series of models can all be explained in similar fashion.

% __________________________________________________

\subsection{Discussion and future work}
\label{subsec:crystdisc}
The goal of this paper is to show how treating the disk as a multidimensional object and correctly solving the problem of sub-Keplerian accretion affect the results of \daw{}. As shown in Figs.\ \ref{fig:cryst} and \ref{fig:cryst3-15}, we obtain smaller fractions of crystalline silicates throughout the disk. This is an improvement over the old model, which was noted to overpredict crystallinity compared with observations.

A detailed parameter study is required to judge how well our current model reproduces all available observations. One complicating factor in such a procedure is the observed lack of correlation between crystallinity and other systemic properties such as the stellar luminosity, the accretion rate and the masses of the star and the disk \citep{watson09a}. The observed absence of a correlation between two observables usually translates to a lack of a physical correlation, but this is not always the case. For example, \citet{kessler07a} showed why the crystallinity is not observed to be correlated with the stellar luminosity. The disk around a brighter protostar is warmer throughout, so the region from which most of the silicate emission originates lies at a larger distance from the star, where the crystallinity is lower. At the same time, though, the higher temperatures mean that more material can be thermally annealed, so the crystallinity at all radii goes up. The two effects cancel each other, so the observed crystalline fraction is not correlated with the stellar luminosity. Likewise, care must be taken when interpreting other observed non-correlations or correlations.

The best starting point for a more detailed comparison between model and observations appears to be the observed radial dependence of the relative abundances of specific types of crystalline silicates, such as enstatite and forsterite. Our model can be expanded to track multiple types of silicates, each with their own formation temperature and mechanism. First of all, this may help in explaining the ``crystallinity paradox'' identified by \citeauthor{olofsson09a} (\citeyear{olofsson09a}; see also Sect.\ \ref{subsec:oldres}). Second, it can address the question whether crystalline silicates are predominantly formed by condensation from hot gas ($\sim$1200 K), by thermal annealing at slightly lower temperatures ($\sim$800 K), or by shock waves outside the hot inner disk. At the moment, neither the observations nor the models can rule out any of these mechanisms. The crystalline fractions obtained with our model suggest that thermal annealing followed by radial mixing must be taking place and must therefore be responsible for part of the observed crystalline silicates. A scenario in which \emph{all} crystalline material is formed where it is observed, according to the model of \citet{bouwman08a}, appears unlikely.

In addition to tracking multiple types of silicates, it may be worthwhile to investigate different collapse scenarios. The \citet{shu77a} collapse starts with a cloud core with an $r^{-2}$ density profile, but observations of pre-stellar cores usually show an $r^{-1.5}$ density profile instead \citep{alves01a,motte01a,harvey03a,andre04a,kandori05a}. Bonnor-Ebert (BE) spheres have such a density profile \citep{ebert55a,bonnor56a}, so they have been proposed as an alternative starting point for collapse models \citep{whitworth96a}. The collapse of a BE sphere results in different densities, velocities and temperatures than those obtained with the \citeauthor{shu77a} collapse \citep{foster93a,matsumoto03a,banerjee04a,walch09a}, leading in turn to different crystalline silicate abundances. However, no analytical solutions currently exist for the collapse of a rotating BE sphere, so we are unable to pursue this point in any more detail.

% ______________________________________________________________________

\section{Conclusions}
\label{sec:conc}
This paper presents a new method of correcting for the sub-Keplerian velocity of envelope material accreting onto an axi\-symmetric two-dimensional circumstellar disk. Unlike the previous corrections of \citet{hueso05a} and \citet{visser09a}, this new method properly conserves angular momentum and produces infall trajectories without discontinuities. The latter is important for tracing changes in the chemical contents and dust properties during the evolution of the envelope and disk.

The disks produced with the new method are smaller than those produced with the old method by up to a factor of ten. Depending on the initial conditions, the disk masses are between 100 and 200\% of previously computed values (Sect.\ \ref{sec:diskprop}). The new disks are a few degrees colder in the inner regions and a few degrees warmer in the outer regions, resulting in lower abundances of CO ice (Sect.\ \ref{sec:gasice}). By the time the system reaches the classical T Tauri stage, at about 1 Myr, the global ice abundances still agree well with observations. Overall, there are no major changes in the gas-ice ratios compared with \citet{visser09a}.

The disk was treated as geometrically flat by \citet{dullemond06a}. As in \citet{visser09a}, we now also take into account the vertical structure when computing the infall trajectories. This results in the bulk of the accretion occuring at larger radii. A smaller fraction of the infalling material now comes close enough to the star to be heated above 800 K, the temperature required for thermal annealing of amorphous silicates into crystalline form. Therefore, the new method produces crystalline abundances that are lower by a few per cent to more than a factor of five compared to the old model. We now obtain a better match with observations and we argue that thermal annealing followed by radial mixing is responsible for at least part of the crystalline silicates observed in disks. An expanded model, which tracks specific forms of crystalline silicate, is required to establish in more detail the importance of this and other possible crystallisation mechanisms.

% ______________________________________________________________________

\begin{acknowledgements}
The authors are grateful to Ewine van Dishoeck for stimulating discussions, and to the anonymous referee for concise and constructive comments on the original manuscript. Astrochemistry in Leiden is supported by a Spinoza Grant from the Netherlands Organization for Scientific Research (NWO) and a NOVA grant.
\end{acknowledgements}

% ______________________________________________________________________

\bibliographystyle{aa}
\bibliography{13604}

\end{document}